\documentclass[prd,preprint,showkeys,showpacs,footinbib,floatfix]{revtex4}
\usepackage{graphicx}
\usepackage{amssymb}
\def\bea{\begin{eqnarray}}
\def\eea{\end{eqnarray}}

\begin{document}
\preprint{RUB-TPII-13/2003}
\title{\hskip4in  
\\
Determining the Proximity of $\gamma^{\ast}N$ Scattering to
the Black Body Limit Using DIS and $J/\psi$ Production}
\date{January 23, 2004}
\author{T. Rogers} 
\email[E-mail:]{rogers@phys.psu.edu}
\affiliation{Department of Physics,
Pennsylvania State University,\\
University Park, PA  16802, USA
}
\author{V. Guzey}
\email[E-mail:]{vadim.guzey@tp2.ruhr-uni-bochum.de}
\affiliation{Institut f{\"u}r Theoretische Physik II,\\
Ruhr-Universit{\"a}t Bochum, \\
D-44780 Bochum, Germany}
\author{M. Strikman}
\email[E-mail:]{strikman@phys.psu.edu}
\affiliation{ Department of Physics,
Pennsylvania State University,\\
University Park, PA  16802, USA}
\author{X. Zu}
\email[E-mail:]{xiaozu@phys.psu.edu}
\affiliation{Department of Physics,
Pennsylvania State University,\\
University Park, PA  16802, USA}
\altaffiliation{Now at Florida State University.}

\begin{abstract}
We use information about DIS and $J/ \psi$ production on hydrogen to model the $t$-dependence of the $\gamma^{\ast} N$
 scattering amplitude.  We investigate the profile function for elastic scattering of hadronic components of the virtual 
photon off both a nucleon and heavy nuclear target, and we estimate the value of the impact parameter where the black body 
limit is reached.   We also estimate the fraction of the cross section that is due to hadronic configurations in the virtual 
photon wave function that approach the unitarity limit.  We extract, from these considerations, approximate lower limits on the 
values of $x$ where the leading twist approximation in DIS is violated.  We observe that the black body limit may be approached within 
HERA kinematics with $Q^{2}$ equal to a few GeV$^2$ and $x \sim 10^{-4}$.  Comparisons are made with earlier predictions by 
Munier {\it et al.}, and the longitudinal structure function is compared with preliminary HERA data.  The principle advantage 
of our method is that we do not rely solely on the $t$-dependence of $\rho$-meson production data.  This allows us to extend 
our analysis down to very small impact parameters and dipole sizes.  Finally, we perform a similar calculation with a $^{208}$Pb target, 
and we demonstrate that the black body limit is already approached at $Q^{2} \sim 20$ GeV$^{2}$ and $x \sim 10^{-4}$.
\end{abstract}

\maketitle

\section{Introduction}
\label{sec:intro}

One of the current theoretical challenges in quantum chromodynamics (QCD) is to describe 
high energy interactions with hadrons in terms of fundamental field theory.  It is observed that high-energy hadron-hadron scattering interactions become completely absorptive (black) at small impact parameters so that elastic scattering can be viewed essentially as a shadow of the inelastic cross section in the sense of Babinet's principle.  If this regime occurs at most of the impact parameters which contribute to the inelastic cross section, then the elastic and inelastic cross sections are equal.  This limit is often referred to as the black body limit (BBL) in analogy with the quantum mechanical situation of scattering from an absorptive share of radius $r$ in which case the total cross section is equal to $2 \pi r^{2}$, (see e.g., problem 1 of section 131 in Ref~.\cite{landau}).  This limit is also loosely referred to as the unitarity limit, although unitarity alone admits cross sections as large as $4 \pi r^{2}$ provided there are no inelastic interactions (see e.g., problem 2 of section 132 in Ref.~\cite{landau}).  In this paper we will assume, in line with experiment, that the amplitude is predominantly imaginary.  In this case, the unitarity limit coincides with the BBL.    
The black body limit in deep inelastic scattering (DIS) is an interesting new regime in QCD where the coupling 
strength is small, but where the leading twist approximation breaks down and new small coupling methods are needed.  
In order to understand the transition to the BBL, it is important to construct models which accurately describe 
$\gamma^{\ast} N$ scattering over a wide range of kinematic variables.  In particular, such a model should interpolate 
smoothly between the nonperturbative domain and the region where interactions are accurately described by leading twist 
perturbative QCD (pQCD).  In a recent paper by Munier {\it et al.} \cite{Munier:2001nr}, an estimate of the proximity to 
the BBL for the interaction of a color dipole with a proton was made using data from diffractive electroproduction of 
$\rho$-mesons \cite{Munier:2001nr}.  The techniques used in \cite{Munier:2001nr} are limited by the need to model the 
$\rho$-meson wave function, as well as by limited information on the $t$-dependence for $\rho$ production.  
Within the impact parameter representation, this means that their predictions are constrained to intermediate values of impact parameter 
($b \gtrsim 0.3$ fm) \cite{confusion}. (It will be  important to keep in mind the distinction between impact parameter, $b$, and dipole size, $d$.)  
We will find that our analysis is valid down to very small values of $b$ and $d$.  
We will make an improved estimate of the onset of the BBL by combining information from DIS and $J/ \psi$ production.  
A great advantage of our technique is that, unlike models which are restricted to using meson production data, our model 
is valid down to very small dipole sizes because we utilize leading twist pQCD for calculations involving small size $q \bar{q}$ pairs.  
Within our model of $\gamma^{\ast} N$ interactions, we always assume that the virtual photon can be written as a linear combination of 
hadronic states.  Furthermore, we relate the hardness of the interaction to the size of the hadronic state as is done in 
Ref.~\cite{McDermott:1999fa}, and we take into account the dependence of the hadronic interaction on the virtuality of the original photon.  Large size configurations constitute the soft component of the interaction whereas small 
size $q \bar{q}$ pairs constitute the hard component.  

Modeling the $\gamma^{\ast} N$ interaction requires making three principle observations.  The first is to recall that 
the total cross section for small size dipole configurations ($d \lesssim .3$ fm) is known within the framework of perturbation theory 
down to $x \sim 10^{-4}$.  Therefore, we will have a complete picture of the interaction of small size configurations with nucleons if 
we can extract the $t$-dependence from experimental data.  In Sect.~\ref{sec:model}, we discuss how the $t$-dependence of the small 
size $q \bar{q}$ pairs can be extracted from $J /\psi$ production data.  Secondly, we note that the soft scattering of large size 
hadronic configurations is well understood phenomenologically in terms of effective Pomeron exchange.  Hence, our model should reproduce 
the pion-nucleon amplitude for hadronic configurations comparable to the size of the pion ($d \approx .6$ fm).  Finally, we must model 
the behavior of the amplitude for intermediate hadronic sizes ($.3$ fm $\lesssim d \lesssim .6$ fm).  This is an interesting and poorly 
understood region of kinematics, and our model will allow for readjustments in the transition region.  
  
In Ref.~\cite{McDermott:1999fa}, a system was devised for relating the transverse size of a $q \bar{q}$ dipole to the virtuality, $\bar{Q}^{2}$.
The pQCD result for the inelastic $\gamma^{\ast} N$ cross section was interpolated to large size hadronic configurations and 
matched to the total cross section for pion-nucleon scattering.  The hadronic size,   
 $d$, is used to interpolate between the hard and soft regions in this paper as it was in Ref.~\cite{McDermott:1999fa}.  
 $d$ represents the transverse size of a quark-antiquark pair in the limit that $d$ is very small ($d \lesssim 0.1$ fm).  
However, as $d$ grows large, the dipole picture becomes inappropriate since the hadronic components that correspond to soft 
interactions consist of large, complex hadronic states.  Larger values of $d$ should be interpreted as transverse sizes of general 
hadronic components of the virtual photon wave function.  The cross section takes the form familiar from perturbation theory:
\begin{eqnarray}
\hat{\sigma}_{pQCD}(d,x) = \frac{\pi^{2}}{3}d^{2} \alpha_{s} (\bar{Q}^{2})x^{'}g(x^{'},\bar{Q}^{2}) \,.
 \label{eq:pqcd} 
\end{eqnarray} 
Here, $\lambda =  \bar{Q}^{2} d^{2}$ is a universal scaling ansatz used to relate energy scales, $Q^{2}$, to transverse dipole sizes, $d$; $x^{'}$ is the light-cone fraction for the gluon attached to the $q \bar{q}$ loop.  
One manifestation of the QCD factorization theorem is that the contribution of hadronic configurations within the photon to the 
 longitudinal cross section, $\sigma_{L}$, is peaked around  a narrow range of small 
 dipole sizes (see Fig.~\ref{fig:dist}).  The value of $\lambda$ is chosen so that, within the perturbative region, 
$d^{2} = \lambda / Q^{2}$
is approximately the average dipole size contributing to $\sigma_{L}$.  For large $Q^{2}$, $\lambda$ 
takes on values of the order of 10.  In fact, it is found that 
$F_2$ and $F_L$ depend very weakly on the value of $\lambda$ within the perturbative region of 
$d$~\cite{McDermott:1999fa}. 
Changing the value of $\lambda$ thereby provides a universal parameter for tuning the cross section within the transition region.  In this paper, we use $\lambda = 4$ because it is found that $\lambda = 4$ best describes $J/ \psi$ 
data over a wide range of kinematics (both perturbative and nonperturbative).  Figure~\ref{fig:Lambda} compares the profile function for $\lambda = 4$ 
and $\lambda = 10$.  The function used in Ref.~\cite{McDermott:1999fa} to interpolate between the hard and soft regions matches 
smoothly to the pQCD result at small $d$ and to the pion-proton cross section at large $d$.  Furthermore, it takes into consideration 
the breakdown of the leading twist formula in the small $x$ limit.

Notice that the $q \bar{q}$ pair is not a fundamental object since it is always produced off-shell by a virtual photon.  This is taken 
into account on the right side of 
Eq.~(\ref{eq:pqcd})
 through 
implicit
 dependence of $\frac{x^{'}}{x}$, on $\bar{Q}^{2}$ and $Q^{2}$
(see Ref.~\cite{McDermott:1999fa} and 
 Eq.~(\ref{eq:xprime})).
 The dependence of the dipole cross section on the external photon virtuality, $Q^{2}$, is a feature that is absent in other models such as the one proposed by W\"{u}sthoff and Golec-Biernat~\cite{Golec-Biernat:1999qd}.  This point will be important for what we discuss later because it means that we cannot speak unambiguously about the dipole cross section with referring to the interaction for which it is a sub-proccess.  

In this paper, we will model the $t$-dependence of the $ \gamma^{\ast} N$ elastic scattering amplitude (which, of course, cannot be observed experimentally) using data from $J/\psi$ photo(electro)-production in conjunction with the pion-proton elastic scattering amplitude. 
 As when we model the behavior of the total cross section in various kinematic regions, we model the $t$-dependence of the $\gamma^{\ast}N$ amplitude by considering three distinct steps. First, we model the $t$-dependence of the small size $q\bar{q}$ configurations.  Because of transverse squeezing in the $J/\psi$ wave function, data taken from $J/\psi$ production is appropriate for use in modeling the $t$-dependence of small size, hard scale $q\bar{q}$ configurations.  Next, the $t$-dependence of soft, large size hadronic configurations is approximated by the pion structure function, the pion being a reasonable approximation to a large size hadronic component of the virtual photon wave function.  For soft interactions, a factor for soft Pomeron exchange is included to account for the slow rise in cross section at small $x$.  Finally, we use the transverse size, $d$, as a parameter to interpolate between hard and 
soft physics.  In our analysis, we will transform our expression for the amplitude into the impact parameter picture to look for the regions where the impact parameter space amplitude approaches the unitarity limit, and thus to estimate the values of the impact parameter where the BBL is attained.

It should be noted that our model has limitations which restrict how it can be applied.  
It is important to keep in mind that Eq.~(\ref{eq:pqcd}) is multiplied by a color factor of 9/4 when the hadronic configuration is a color singlet composed of octet representations of SU(3), as in a gluon dipole.  Such interactions are expected to be abundant in the small $x$ limit, so the BBL will be reached at larger impact parameters than what is predicted by considering only the interactions of small size $q \bar{q}$ pairs.  Furthermore, at very small $x$, effects from cross section fluctuations of the virtual photon become important, and taking into account only elastic dipole scattering becomes inappropriate.  In particular, the total cross section in hadron-hadron scattering has a significant contribution from inelastic diffraction.  In Ref.~\cite{Miettinen:1978xv}, Miettinen and Pumplin write the contribution from inelastic diffraction in terms of fluctuations around hadronic eigenstates.
Estimates of the contribution from hadronic fluctuations
to the cross section show that it is not negligible  
(see, e.g. Ref.~\cite{Blaettel:ah}).  In fact, with a decrease of the dipole size, the relative importance of inelastic diffraction increases.  The importance of inelastic diffraction for small $q\bar{q}$ sizes is discussed, for example, in Ref.~\cite{Frankfurt:2002kd}.
 The important point to note here is that inelastic diffraction will contribute to the breakdown of the leading twist approximation at low $x$ before the BBL for elastic scattering of the hadronic configurations in the photon wave function is reached.  As such, we do not seek to place an absolute boundary on the region where corrections to the DGLAP evolution equation become relevant. 
Rather, we construct a model that puts a lower limit in impact parameter space on regions approaching the BBL.  Furthermore, since the leading twist approximation is not accurate in the vicinity of the unitarity limit, the BBL establishes a lower limit in impact parameter space on regions where the DGLAP equation is applicable.    

In 
Sect.~\ref{sec:model},
 we outline our model for the $t$-dependence of the hadronic configuration-nucleon amplitude.  
In Sect.~\ref{sec:impact}, we transform our expression for the amplitude into the impact parameter representation and 
study the proximity of the profile function to the unitarity limit as a function of impact parameter, $b$.  
We write the transverse and longitudinal cross sections, 
$\sigma^{ \gamma^{\ast} N}_{T,L}$, as convolutions of the dipole cross section with the photon wave function
in Sect.~\ref{sec:gamma}.
  Section ~\ref{sec:gamma} concludes with a calculation of the fraction of the $\gamma^{\ast}N$ cross section due to large values of the hadronic profile function.  In 
Sect.~\ref{sec:compare}, we compare our results to an earlier study of the S-matrix $t$-dependence in impact parameter space.  In Sect.~\ref{sec:nuclear} we perform the same calculation for the situation where the target is a $^{208}$Pb nucleus.  For the case of a nuclear target, we match the pQCD calculation at small $d$ to the Glauber multiple scattering theory result at large $d$. Finally, we summarize our observations in the conclusion.

\section{Modeling the $t$-Dependence}
\label{sec:model}

Starting with the expression for the total cross section in Eq.~(\ref{eq:pqcd}), we devise a model for the scattering amplitude by writing it in the form, 
\begin{eqnarray}
A_{hN}(s,t) = is \hat{\sigma}_{tot} f(s,t) \,,
\end{eqnarray}
where $f(s,t)$ accounts for the $t$-dependence of the interaction, and $\hat{\sigma}_{tot}$ is determined from the QCD improved dipole picture (Eq.~(\ref{eq:pqcd})).  The ``hat'' on $\hat{\sigma}_{tot}$ is to distinguish the total cross section for the scattering of one component of the photon wave function from the total $\gamma^{\ast}N$ cross section which we consider in Sect.~\ref{sec:gamma}.  Applying the optical theorem in the large $s$ limit reproduces $\hat{ \sigma }_{tot}$.  For now we assume that the amplitude is purely imaginary.  We will return to the question of a real part of the amplitude at the end of Sect.~\ref{sec:impact}.

In this section, we will make an estimate of the form of $f(s,t)$ which will take into account the nonzero size of the $q \bar{q}$ dipole and which will smoothly interpolate between perturbative and nonperturbative regimes using the hadronic size, $d$, as a parameter.  We  act in the spirit of 
Ref.~\cite{McDermott:1999fa} by modeling the $t$-dependence of the amplitude in the soft and hard regimes and by using $d$ to build a smooth interpolation.  The three steps:  building a model for the small dipole region, building a model for the large wave-packet region, and interpolating between the two regions are outlined in the next three paragraphs.   

We start by writing the general structure of the amplitude.  The $t$-dependence, $f(t,x,d)$, is written as the product of three functions 
\begin{eqnarray}
f(t,x,d) = F_{N}(t,d) F_{h}(t,d) F_{P}(t,x,d) \,.        
\label{eq:largesize}
\end{eqnarray}
Here and in the rest of this section, the dependence of $f$ upon $s$ is replaced by dependence upon $x$ and $d$. 
$F_{N}(t,d)$ describes the $t$-dependence of the nucleon target, $F_{h}(t,d)$ describes the $t$-dependence of the hadronic projectile, and $F_{P}(t,x,d)$ accounts for Gribov diffusion.  This method of separating the $t$-dependence into three factors corresponding to different sources of $t$-dependence is similar to what is used in Ref.~\cite{Povh:ju}. 
 
The next task is to model the small dipole size $t$-dependence.  Both the soft Pomeron exchange factor and the hadronic form factor must approach unity as the size, $d$,
 shrinks to zero.  The QCD factorization theorem implies that the $t$-dependence of the small dipole-nucleon amplitude is universal.  Hence, it can be extracted directly from $J/ \psi$ photo(electro)-production since the $J/ \psi$ wave function is known to be a small size wave-packet~\cite{Frankfurt:1996ri}.  Data from $J/ \psi$ production reveal that the two-gluon 
form factor is,
\begin{eqnarray}
F_{N}(t,d \rightarrow 0) = F_{1}(t) \sim \frac{1}{ (1 - t/m_{1}^{2})^{2}} \,. \label{eq:gluon}
\end{eqnarray} 
The subscript, 1, labels the two-gluon form factor and $m_{1}^{2}$ is a measurable parameter in the two-gluon form factor.  The value, $m_{1}^{2} \approx 1.1$ GeV$^{2}$ is extracted from data in Refs.~\cite{Levy:1997bh,Adloff:2000vm,Chekanov:2002xi}. 
For a detailed discussion of the two-gluon form factor in $J/\psi$ production, 
see Ref.~\cite{Frankfurt:2002ka}.
 In \cite{Frankfurt:2002ka}, it was discussed in detail how, due to the transverse squeezing of the $J/ \psi$ wave function, the $t$-dependence of the amplitude comes solely from the two-gluon form factor.  
In particular, it was found 
 that the dipole form factor contributes only about $0.3$ GeV$^{-2}$ to the slope of the 
$t$-dependence.  The assumption that only the gluon form factor is relevant for $J/ \psi$ production has been successfully tested against data in Refs.~\cite{Binkley:1981kv,Camerini:1975cy,Gittelman:1975ix}.   Hence, in the limit of small dipole sizes,
\begin{eqnarray}
f(t,x,d \rightarrow 0) = F_{1}(t) \label{eq:smallsize} \,.
\end{eqnarray}      

Next we construct a model for the large wave packet behavior.  When the hadronic state has a large size, the $t$-dependence receives contributions from sources other than the two-gluon form factor.  We rewrite Eq.~(\ref{eq:largesize}) in the form,
\begin{eqnarray}
f(t,x,d) = F_{N}^{e.m.}(t,d)F_{h}(t,d)F_{P}(t,x,d).
\end{eqnarray}
Now, $F_{N}^{e.m}(t)$ is the electromagnetic form factor of the nucleon which is known phenomenologically to take the
form,
\begin{eqnarray}
F^{e.m.}_{N}(t) \sim \frac{1}{ (1 - t/m_{0}^{2})^{2}} \,,
 \label{eq:nucleon}
\end{eqnarray}
where $m_{0}^{2} \approx 0.7$ GeV$^{2}$. Large size hadronic configurations can be reasonably expected to have $t$-dependence similar to the pion
electromagnetic form factor.  Thus, for the hadronic form factor
 we use the well-known form of the pion form factor,
\begin{eqnarray}
F_{h}(t,d \rightarrow d_{\pi}) \sim \frac{1}{1 - t/m_{2}^{2}} \,, \label{eq:pion}
\end{eqnarray}
with $m_{2}^{2} \approx 0.6$  GeV$^{2}$.  
Here, $d_{\pi}$ is the characteristic size of the pion, and takes on a value of approximately $0.65$ fm.  This value for the pion size is consistent with what is used in the matching ansatz of Ref.~\cite{McDermott:1999fa} and agrees well with data for the $\pi$N cross section in Ref.~\cite{Burq:1981nv}. 
  For low-$x$
 soft scattering there is also a factor that arises from Gribov diffusion effects:
\begin{eqnarray}
F_{P}(t,x,d \rightarrow d_{\pi}) \sim e^{- \alpha^{\prime} t \ln \frac{x_{0}}{x}} \,. \label{eq:diff}
\end{eqnarray}
The factor, $F_{P}(t,x,d)$, describes the exchange of a soft Pomeron with Regge slope $\alpha^{\prime} \approx 0.25$ GeV$^{-2}$ and $x_{0} = 0.01$.  The value of $x_{0}$ is determined by the boundary of the region where Gribov diffusion effects become significant.  

Finally, we must find a reasonable way to interpolate between the hard and the soft regions.  We use the $t$-dependence discussed in the previous two paragraphs to guess the following form for the hadronic configuration-nucleon amplitude:
\begin{widetext}
\begin{eqnarray}
A_{hN}(s,t) = i s \hat{\sigma}_{tot} \frac{1}{ (1 - t/M^{2}(d^{2}))^{2}} \frac{1}{1 - t d^{2} /d_{\pi}^{2} m_{2}^{2}} e^{ \alpha^{'} \frac{d^{2} t}{d_{\pi}^{2}} \ln \frac{x_{0}}{x}} \label{eq:ourmodel} \,.
\end{eqnarray}
\end{widetext}    
In order to give the variation with $d$ geometric behavior, we use $d^{2}$ as a parameter.  To interpolate between the nucleon and the two-gluon form factors, we have defined the function,
\begin{eqnarray}
M^{2}(d^{2}) = \left\{ \begin{array}{ll}
             m_{1}^{2} - (m_{1}^{2} - m_{0}^{2})\frac{d^{2}}{d_{\pi}^{2}} & ,\mbox{\hspace{2mm}$d \leq d_{\pi}$} \\
                                                               m_{0}^{2} & ,\mbox{\hspace{2mm}otherwise}         
                       \end{array} \right. \,. \label{eq:M}
\end{eqnarray}
Note that when $d$ equals $d_{\pi}$, $A_{hN}$ is the product of Eqs.~(\ref{eq:nucleon}), 
(\ref{eq:pion}), and (\ref{eq:diff}).
  In the small $d$ limit, the dipole form factor and the Pomeron form factor approach 
unity,
 $M^{2}(d^{2}) \rightarrow m^{2}_{1}$, and the limit in 
Eq.~(\ref{eq:smallsize})
 is recovered.  Varying $d^{2}$ interpolates smoothly between the hard and soft regions.  Note that we neglect a possible small $x$ dependence of $F_{N}(t,d)$ at $x \lesssim 0.01$.  (See the discussion in Ref.~\cite{Strikman:2003gz}.)  However, our model is adjusted to reproduce the observed $x$-dependence of the slope for photoproduction of $J/ \psi$ mesons.  

\section{Impact Parameter Analysis}
\label{sec:impact}

\begin{figure*}
\rotatebox{270}{\includegraphics[scale=0.60]{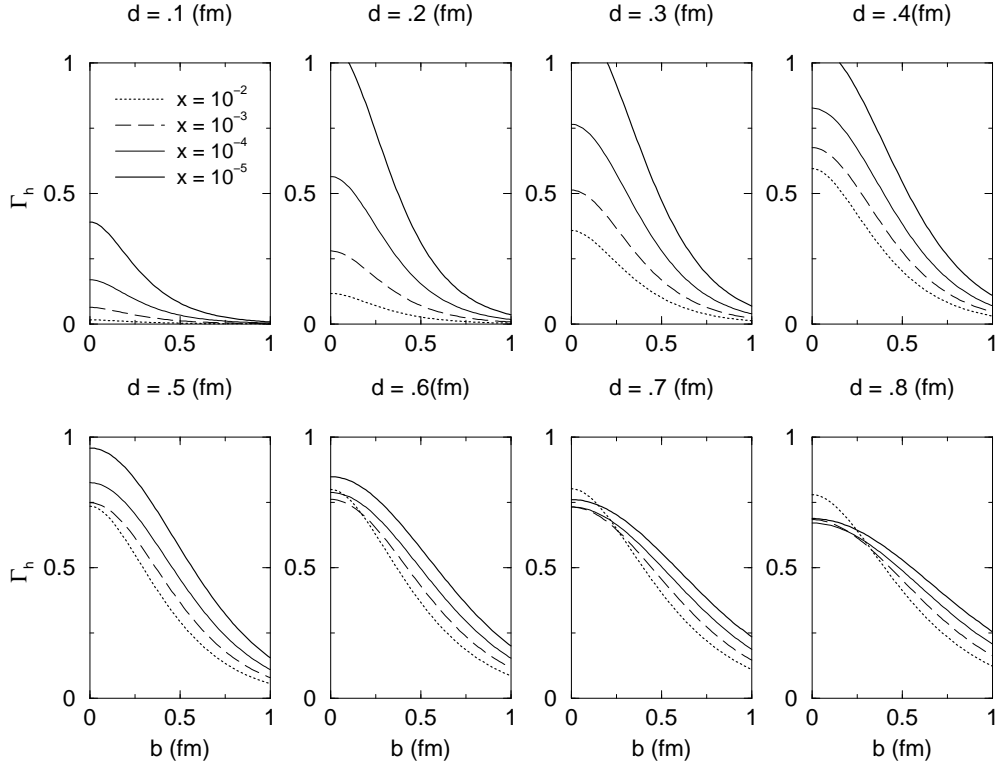}}
\caption{\label{fig:dipoleg}The hadronic configuration-nucleon profile function for different $x$ values.  The large $\Gamma_{h}(b)$ region (\
$\Gamma_{h} \gtrsim 1/2$) is reached for intermediate hadronic sizes.  (See figure \ref{fig:fraction}.)  Here, $Q^{2}$ is taken to be 2 GeV$^2$.}
\end{figure*}

Having obtained Eq.~(\ref{eq:ourmodel}), the next step is to transform to the impact parameter representation where the profile function is defined by the relation,
\begin{eqnarray}
A_{dipole, N}(s,t) = 2 i s \int d^{2} \vec{b} e^{ - i \vec{q} \cdot \vec{b} } \Gamma_{h}(s,b) \,, \label{eq:amplitude}
\end{eqnarray}
where
 $t = -q^{2}$.  The subscript, $h$, indicates that we are considering the profile function for the scattering of a single hadronic component of the photon wave function from the proton.  We get the profile function by inverting Eq.~(\ref{eq:amplitude}),     
\begin{eqnarray}
\Gamma_{h}(s,b) = \frac{1}{2is(2 \pi)^{2}} \int d^{2} \vec{q}  e^{i \vec{q} \cdot \vec{b}} A_{hN}(s,t) \,.
 \label{eq:invamp}
\end{eqnarray}          
For an imaginary amplitude, the BBL is reached when $\Gamma_{h}(s,b) = 1$ and the elastic and inelastic cross sections are equal.
Recall that if a singlet dipole consists of color octet representations of SU(3),
 Eq.~(\ref{eq:pqcd}) has an extra factor of $9/4$, so the BBL for the interaction of a hadronic configuration with the nucleon is certainly reached for $\Gamma_{h}(b) \sim 1/2$.  Therefore, whenever $\Gamma_{h} \gtrsim 1/2$, we conclude that the interaction takes place near the BBL.     
In the rest of this section, we will suppress explicit reference to the argument, $s$, in the profile function.  We have plotted
 the function $\Gamma_{h}(b)$ for different values of the dipole size and $x$ in    
Fig.~\ref{fig:dipoleg}.  We have used gluon distributions from CTEQ5L in the perturbative calculation of $\hat{\sigma}_{tot}$ \cite{Lai:1999wy}.  Recall from the introduction that our model requires that we specify the external photon virtuality.  Since we are interested in the possibility of reaching the BBL at a few GeV$^2$, we have set $Q^{2} = 2$ GeV$^{2}$ in Fig.~\ref{fig:dipoleg}.   

\begin{figure}
\rotatebox{270}{\includegraphics[scale=0.60]{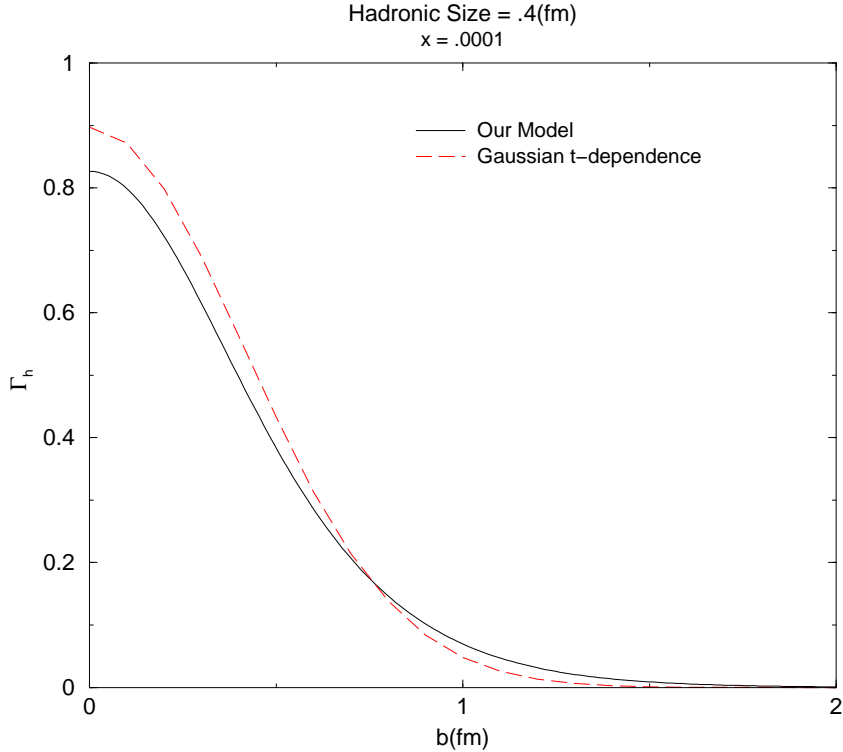}}
\caption{\label{fig:compgauss}Comparison of the $b$ behavior for our model with that of a Gaussian model.  Our model falls off more slowly with $b$.  The slope of the Gaussian used here is $0.17$ fm$^{2}$.}
\end{figure}

A Gaussian
ansatz
 is commonly used in experiments to extrapolate the $t$-dependence to large values.  Let us, therefore, briefly compare the behavior of our model to that of a simple Gaussian.  
  We start with the form
\begin{eqnarray}
A_{hN}(s,t) = is \hat{\sigma}_{tot} e^{C t /2} \,, 
\end{eqnarray}
which
 is then transformed into impact parameter space giving,
\begin{eqnarray}
\Gamma_{hN}(b) = \frac{\hat{\sigma}_{tot}}{4 \pi C} e^{\frac{-b^{2}}{2 C}} \,.
\end{eqnarray}
The slope of the Gaussian, $C$, is chosen so that it yields the same standard deviation in $\Gamma_{h}(b)$ as our model.
One danger in using a Gaussian model is that it neglects the importance of
 interactions in peripheral regions.  Our model attempts to fix this problem by spreading out the distribution in $t$.  Note in Fig.~\ref{fig:compgauss} that our model fall off more slowly with $b$. 

\begin{figure*}
\rotatebox{270}{\includegraphics[scale=0.60]{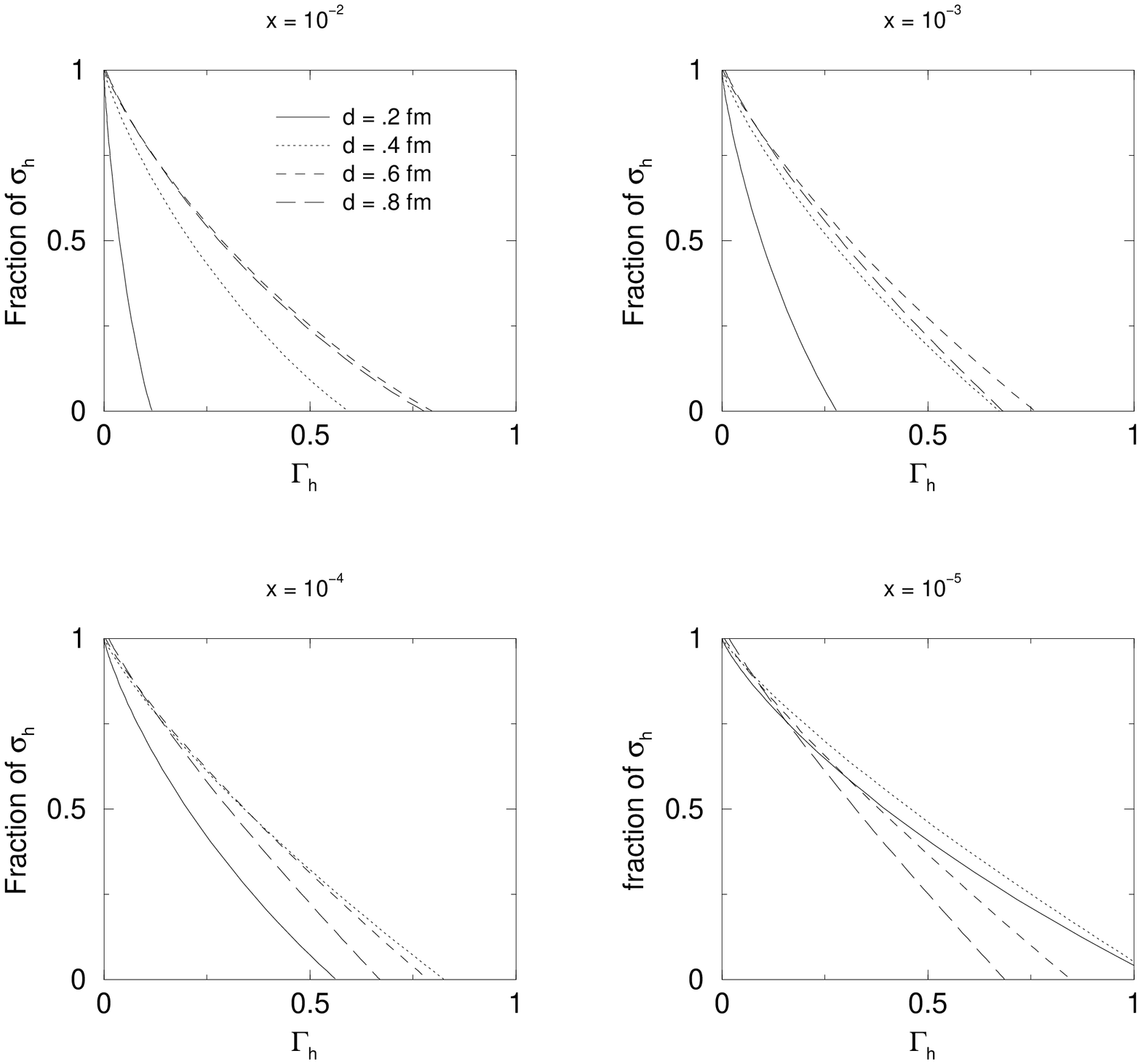}}
\caption{\label{fig:fraction}The fraction of $\hat{\sigma}_{tot}$ with contributions coming from values of $\Gamma_{h}$ greater than the corresponding point on the $x$-axis.}
\end{figure*}

\begin{figure}
\rotatebox{270}{\includegraphics[scale=0.60]{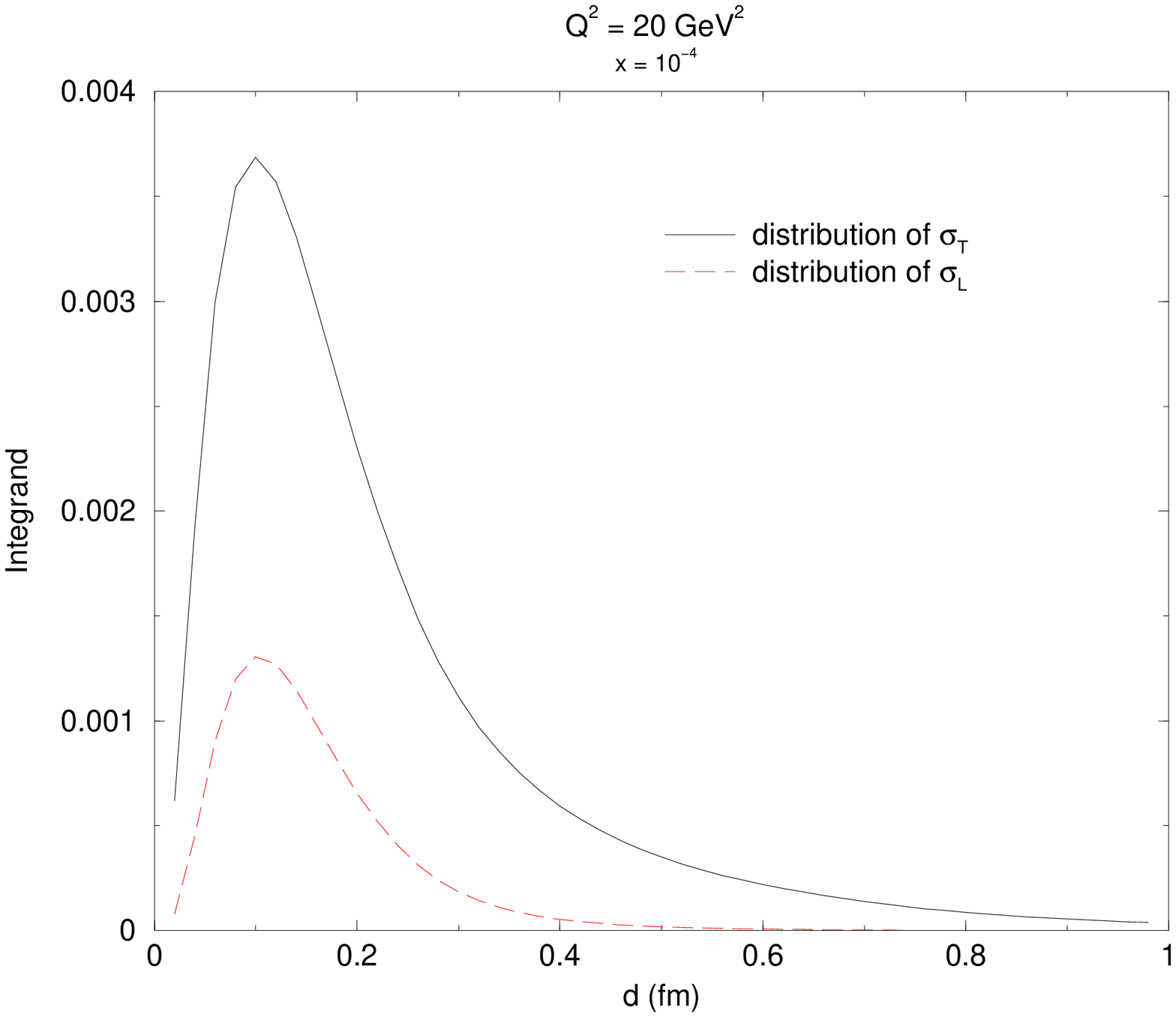}}
\caption{\label{fig:dist}The distribution of the integrand in Eq.~(\ref{eq:totalcross}) over hadronic sizes for both the transverse and longitudinal cross sections.}
\end{figure}

Now let us estimate the contribution of large values of $\Gamma_{h}(b)$ to the total hadronic cross section.  The total cross section follows from the optical theorem,
\begin{eqnarray}
\hat{\sigma}_{tot} = 2 \int d^{2} \vec{b} Re{\Gamma}(s,b). \label{eq:cross} 
\end{eqnarray}
We have made a numerical evaluation of the fraction of the total hadronic configuration-nucleon cross 
section obtained by setting different upper limits on the $b$-integral in Eq.~(\ref{eq:cross}).  
In Fig.~\ref{fig:fraction}, one can see that no more that about thirty percent of the total hadronic cross section is due to values of $\Gamma_{h} \gtrsim 1/2$.  Moreover, contributions from large values of $\Gamma_{h}(b)$ occur for hadronic sizes close to the pion size, $d \approx .6$ fm.  Averaging over the photon wave function will lead to a suppression of contributions from larger size hadronic configurations, so there will indeed be a small contribution to the total DIS cross section due to large values of $\Gamma_{h}(b)$ (see Fig.~\ref{fig:dist}).  The goal of Sect.~\ref{sec:gamma} will be to determine whether the contribution to the $\gamma^{\ast}N$ cross section from large values of $\Gamma_{h}(b)$ is significant enough that we may expect to see black body behavior within HERA kinematics. 
  
To summarize, Fig.~\ref{fig:dipoleg} demonstrates that large values of $\Gamma_{h}(b)$  
are approached for hadronic configuration-nucleon scattering at central impact parameters, $b \lesssim 0.5$ fm.  In Fig.~\ref{fig:dipoleg} it is seen that this is particularly true for hadronic sizes around $d \approx 0.6$ fm.  Figure ~\ref{fig:fraction} shows that for $d \sim 0.6$ fm, a maximum of about 1/3 of the total hadron-nucleon cross section comes from values of $\Gamma_{h}(b)$ that approach the black limit.  When $d \lesssim 0.2$ fm, a very small fraction of the total hadronic configuration-nucleon cross section comes from large values of $\Gamma_{h}(b)$.  The only contribution from $\Gamma_{h} \gtrsim 1/2$ to the total cross section for $d \lesssim .2$ fm occurs at very small $x$ ($x \lesssim 10^{-4}$).

\begin{figure}
\rotatebox{270}{\includegraphics[scale=0.60]{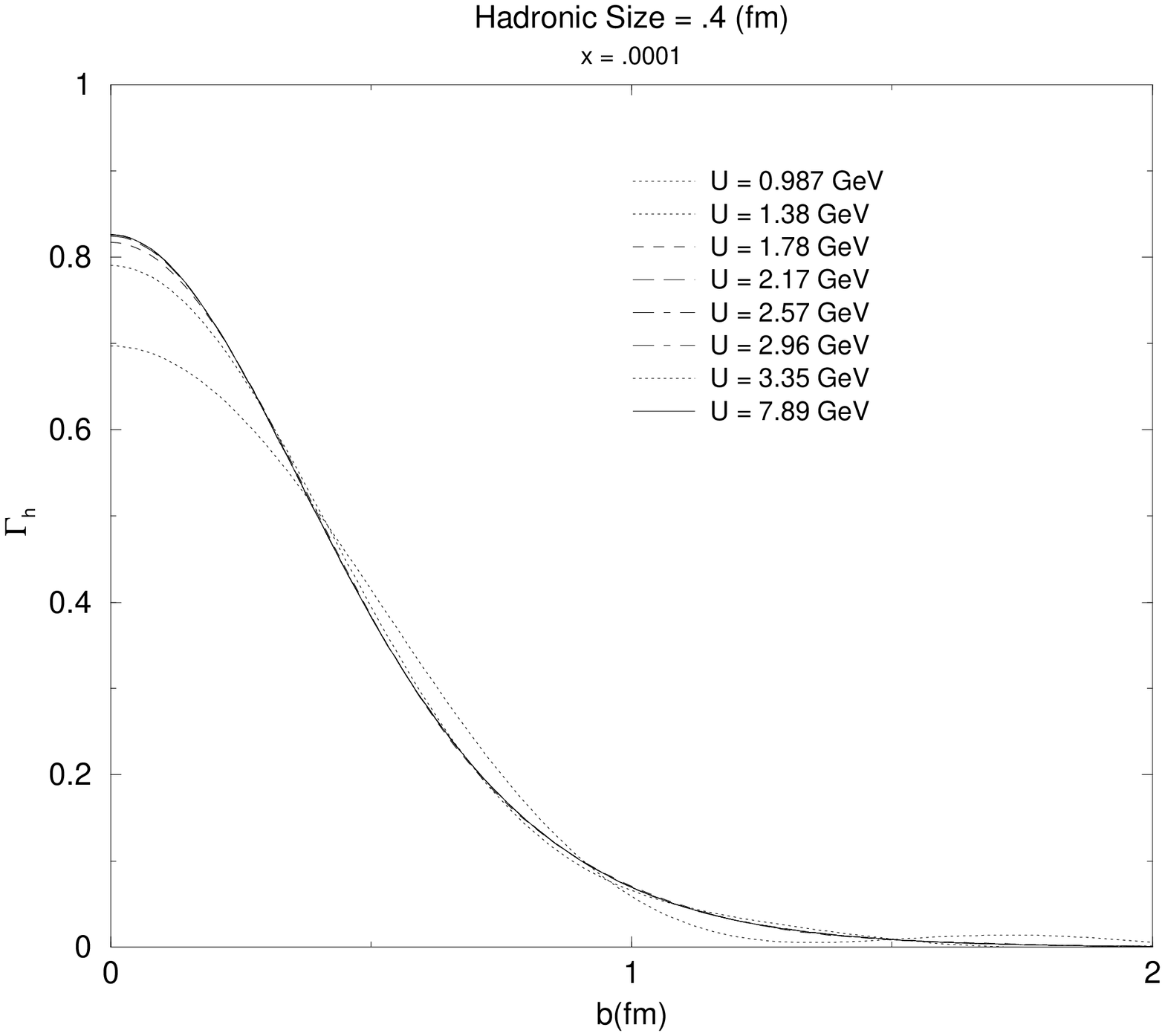}}
\caption{\label{fig:testmodel}A demonstration of the rapid convergence of the profile function.  Here, the profile function is plotted for different values of the upper limit, U, on the integral over t ($-t = U^{2}$).}
\end{figure}

Most of the model dependence in this calculation comes from uncertainty in the large-$t$ behavior of the amplitude.  The different curves in  Fig.~\ref{fig:testmodel} demonstrate how our model changes if we remove contributions from large $t$.  Notice that simply removing the contribution from $-t \gtrsim 3.3$ GeV$^2$ leads to an error of less than ten percent.  Thus, we do not expect our uncertainty in the large $t$ behavior to have a drastic effect.  

We should also remark that we have considered only the non-spin-flip interactions.  Corrections which account for the spin-flip amplitude would result in a smaller non-spin-flip amplitude than what we consider here.  Experimental results in Ref.~\cite{Auer:1977cr} demonstrate that the polarization, $P$, is less than $0.2$ for the range of $t$ we are discussing.  From the formula relating $P$ to the spin-flip amplitude,
\begin{eqnarray}
P = - \frac{2 Im(A_{++}A^{\ast}_{+-})}{|A_{++}|^{2} + |A_{+-}|^{2}} \approx - \frac{2 |A_{+-}|}{|A_{++}|},
\end{eqnarray}
we find the fraction, $\frac{|A_{+-}|}{|A_{++}|} \lesssim 0.1$.  Here, $A_{++}$ represents the amplitude with no spin-flip whereas $A_{+-}$ represents the amplitude with spin-flip. 

We now return to the issue of a real part of the amplitude which we ignored in Sect.~\ref{sec:model}.  In the considered kinematic region ($t \lesssim -2$ GeV$^2$ ), the ratio of the real to imaginary part of the amplitude, $\eta$, is rather small.  Indeed, if we adopt power law behavior for the total cross section, $\hat{\sigma}_{tot} \sim s^{\rho}$, we can estimate the value of $\eta(0)$ using the following formula which follows from the Gribov-Migdal result~\cite{Gribov:1968uy} at high energies in the near forward direction,  
\begin{eqnarray}
\eta(t) = \frac{ReA_{hN}(s,t)}{ImA_{hN}(s,t)} = \frac{\pi}{2} \frac{\partial \ln \hat{\sigma}_{tot}}{ \partial \ln s} = \frac{\pi}{2} \rho. \label{eq:disp}
\end{eqnarray}
The amplitude can be rewritten as,
\begin{eqnarray}
A_{hN}(s,t) \rightarrow s(i + \eta(t)) f(s,t), \label{eq:imaginamp}
\end{eqnarray} 
where the function, $f(s,t)$, is assumed to be strictly real.
For soft regions, the total cross section has the approximate s-behavior of the $\pi N$ cross section as in Ref.~\cite{McDermott:1999fa}, consistent with the behavior of a Donnachie-Landshoff soft Pomeron~\cite{Donnachie:xh}.  In that case, $\rho \approx 0.08$, and Eq.~(\ref{eq:disp}) gives $\eta \approx 0.1$.  The second term in Eq.~(\ref{eq:imaginamp}) appears squared in the calculation of the cross section, so the correction to the cross section is approximately one percent.  For the high $Q^{2}$, low-$x$ region,  the total cross section experiences rapid growth and $\rho \approx 0.25$, or, by Eq.~(\ref{eq:disp}), $\eta \approx 0.35$.  The correction to the squared amplitude is therefore approximately ten percent near the forward direction.  Away from the forward direction, one must account for the small variation of $\eta$ with $t$.    
The effect can be estimated by considering the signature factor in the general form of the Reggeon amplitude.  For $-t \leq 2.0$ GeV$^{2}$, $\eta(t)$ continues to contribute a negligible amount to the amplitude.  

The elastic cross section associated with Eq.~(\ref{eq:imaginamp}) is found by integrating the profile function over impact parameters,
\begin{eqnarray}
\hat{\sigma}_{el} = \int d^{2} \vec{b} | \Gamma (s,b) |^{2}. \label{eq:elcross}
\end{eqnarray}
The total cross section is found using Eq.~(\ref{eq:cross}).
Therefore, the inelastic cross section is,
\begin{eqnarray}
\hat{\sigma}_{inel} = \int d^{2} \vec{b} ( 2 Re \Gamma (s,b) - | \Gamma(s,b) |^{2} ), \label{eq:inelcross}
\end{eqnarray}
with the unitarity constraint,
\begin{eqnarray}
2 Re \Gamma (s,b) - | \Gamma(s,b) |^{2}  \leq 1. \label{eq:unitconst}
\end{eqnarray}  
If the amplitude is purely imaginary, then $\hat{\sigma}_{el} \leq \hat{\sigma}_{inel}$ and the unitarity constraint is that $\Gamma_{h} \leq 1$.
Note that by considering only the imaginary part of the amplitude, we have considered only the real part of the profile function.
If the amplitude is given a real part correction, then the profile function will obtain an imaginary part, the elastic cross section will increase, and the inelastic cross section will decrease.  
The correction to the unitarity constraint on $Re\Gamma$ is, from Eq.~(\ref{eq:unitconst}), $-(\eta Re( \Gamma(s,b))^{2}$.  
In the region of large $Q^{2}$, the effect of a real part in the amplitude would clearly be noticeable.  By Eq.~(\ref{eq:unitconst}), the unitarity limit on the real part of the profile function for $\eta \approx 0.35$ would be,
\begin{eqnarray}
Re\Gamma \leq \frac{1}{1 + \eta^{2}} \sim 0.9. 
\end{eqnarray}
     
\begin{figure}
\rotatebox{270}{\includegraphics[scale=0.60]{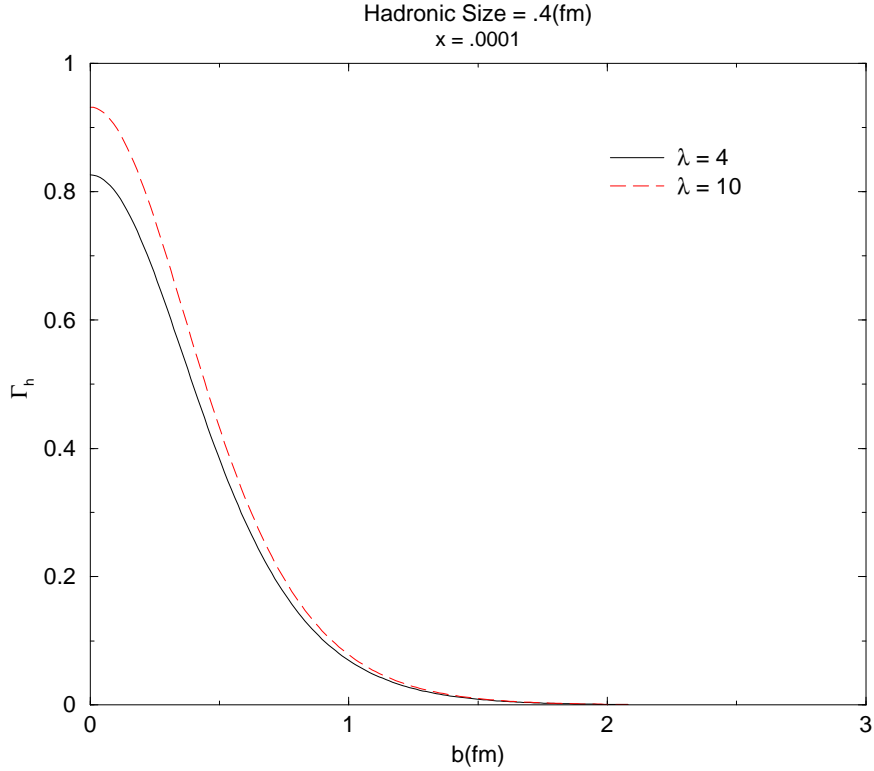}}
\caption{\label{fig:Lambda}Comparison of the profile function for different values of $\lambda$ in the case of an intermediate hadronic size \
equal to $0.4$ fm.  $\Gamma_{h}$ changes by about fifteen percent at $b = 0$ if $\lambda$ is changed from 4 to 10.}
\end{figure}

Thus, the unitarity limit on the real part of the profile function may be less than unity by as much as ten percent.  At this point, we should remark that both the contribution from the real part of the amplitude and the contribution from inelastic diffraction will tend to \emph{raise} the boundary in impact parameter space where the BBL is reached.  We neglect both effects in our model.  As a consequence, when our model predicts that the BBL has been reached below a certain impact parameter,  we can be confident that the same would be true in a model that incorporates inelastic diffraction and the effects of a real component of the amplitude.  On the other hand, if our model predicts that the BBL has not been reached, we must keep in mind that corrections due to inelastic diffraction and a real part of the amplitude may be important.  In other words, the BBL may already be approached at larger values of $b$ than what our model predicts.  

\begin{figure*}
\rotatebox{270}{\includegraphics[scale=0.60]{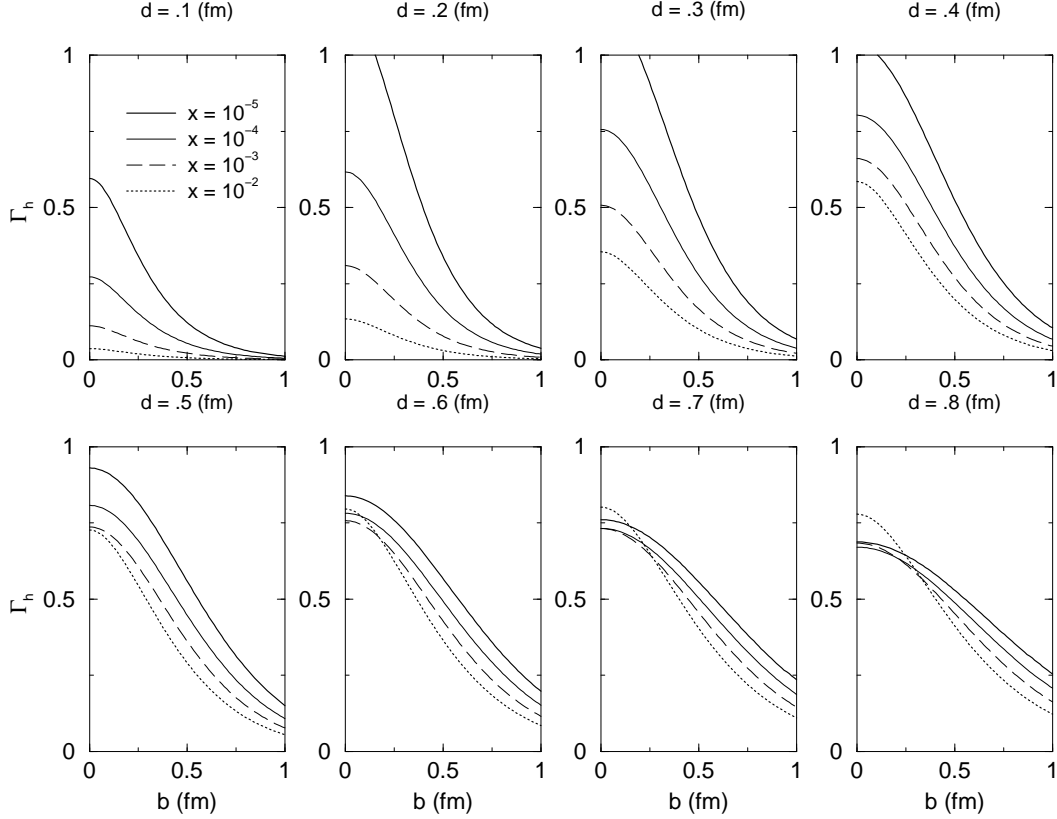}}
\caption{\label{fig:profile_9_23}These graphs are identical to those in Fig.~\ref{fig:dipoleg} except that the value of $Q^{2}$ used to make each graph is calculated from the hadronic size.  Note the larger values of the profile function at small $d$ compared with Fig.~\ref{fig:dipoleg}.}
\end{figure*}

\begin{figure*}
\rotatebox{270}{\includegraphics[scale=0.60]{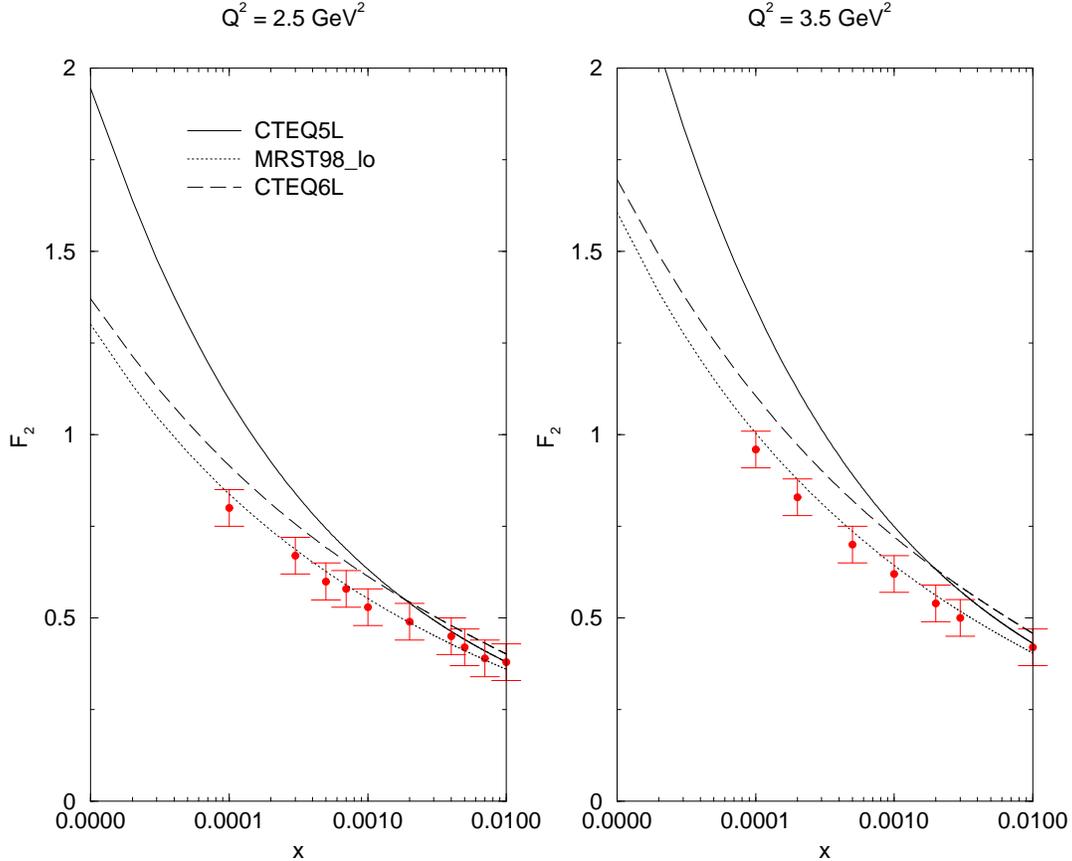}}
\caption{\label{fig:F2vsx}Demonstration of reasonable agreement between the color dipole model and recent HERA data \cite{Lobodzinska} for $F_2$ at low $Q^{2}$.  The different curves correspond to the different parton distributions CTEQ6L,CTEQ5L, and MRST98~\cite{Pumplin:2002vw,Lai:1999wy,Martin:1998sq}.  In our calculations we used CTEQ5L parton distributions because this yields optimal agreement between the dipole model and current data.}
\end{figure*}

\begin{figure*}
\rotatebox{270}{\includegraphics[scale=0.60]{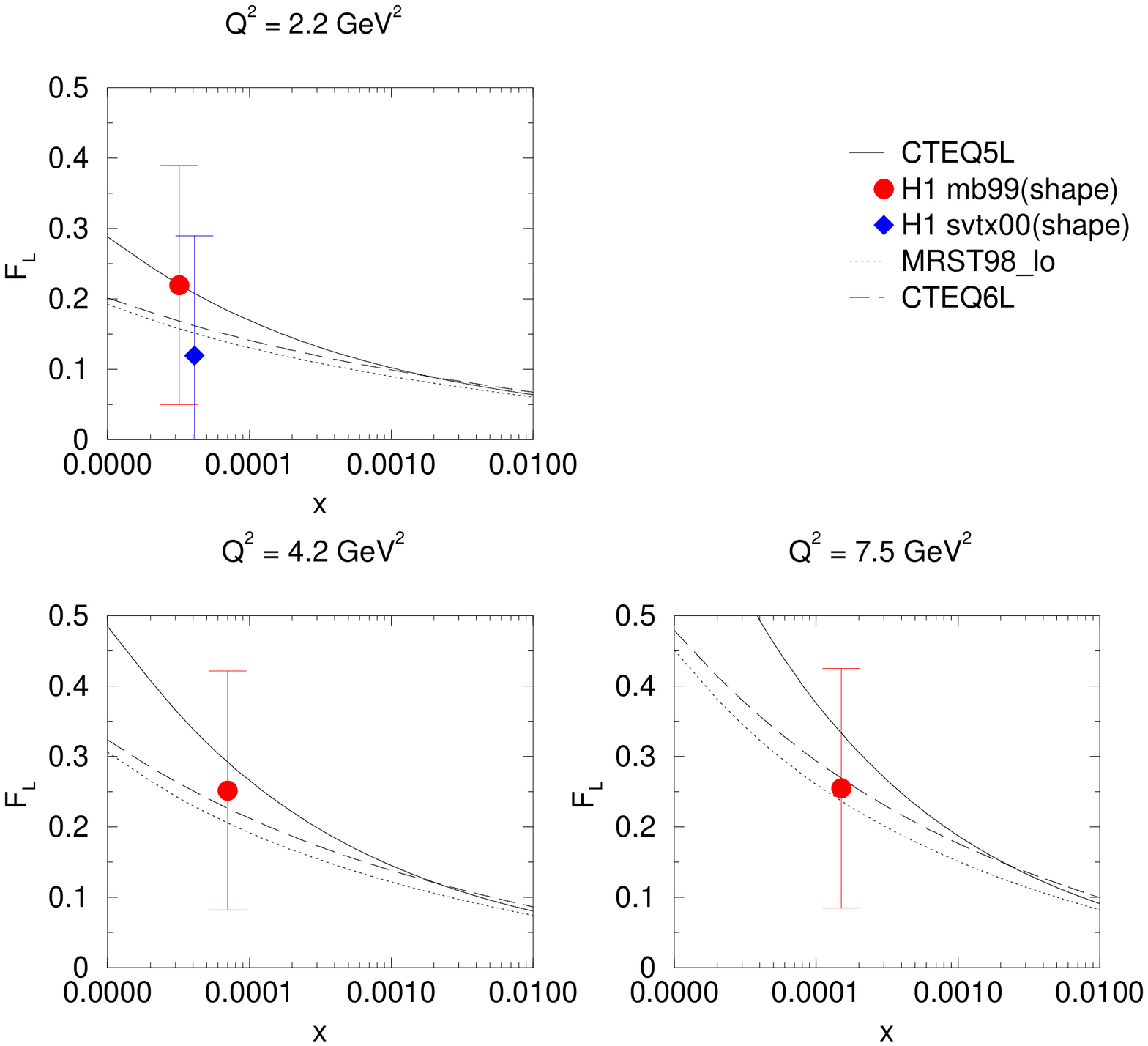}}
\caption{\label{fig:Flvsx}Demonstration of reasonable comparison between the color dipole model and preliminary HERA results \cite{Lobodzinska} for $F_L$ at low $Q^{2}$ and low $x$.  The two points in the upper graph correspond to different methods of taking data.  The different curves correspond to different parton distributions, CTEQ6L,CTEQ5L, and MRST98~\cite{Pumplin:2002vw,Lai:1999wy,Martin:1998sq}.}
\end{figure*}

The uncertainty in the matching region is expressed by the uncertainty in the parameter, $\lambda$.  However, values of the order of 4 to 10 seem to work well and, as shown in Fig.~\ref{fig:Lambda}, there is a variation of only about fifteen percent at small impact parameters when we vary $\lambda$ from 10 to 4.  Note that this is done for a hadronic size of $0.4$ fm which is in the region where the dependence upon $\lambda$ should be at its greatest.  However, there remains another subtlety related to the matching of kinematic regions.  First, recall the distinction between the energy scale, $Q^{2}$, denoting the virtuality of the photon in a particular scattering process, and the scale, $\bar{Q}^{2}$, which is the energy scale related to the hadronic size, $d$, through the scaling ansatz of Ref.~\cite{McDermott:1999fa}.  These two scales are nearly equal as long as we consider hadronic sizes in the vicinity of the average hadronic size for $F_2$.  In determining which value of $x$ (called $x^{'}$ in 
Eq.~(\ref{eq:pqcd}))
 should be used to calculate the hadronic cross section, the authors of Ref.~\cite{McDermott:1999fa} chose to relate the value of $x^{'}$ to the value $x$ for a particular $\gamma^{\ast}N$ process in such a way that $x^{'}$ varies as $d^{-2}$ and so that for typical hadronic sizes, $\langle x \rangle = x^{'}$.  Within the color dipole picture, this accounts for the dependence of the the light-cone fraction, $x^{'}$, of the gluon attached to the $q\bar{q}$ pair on the mass, $M^{2}$, of the dipole:  
\begin{eqnarray}
x^{'} = \frac{Q^{2} + M^{2}}{s}.
\end{eqnarray}
The result is that we cannot speak unambiguously of the hadronic cross section without referring to the virtuality of the probe which generated a component of given transverse size.  From Ref.~\cite{McDermott:1999fa} we have, neglecting the constituent quark mass,
\begin{eqnarray}
x^{'} = x (1 + 0.75 \frac{\bar{Q}^{2}}{Q^{2}}).
\label{eq:xprime}
\end{eqnarray} 
Here we see that if we consider a fixed $Q^{2}$, the universality of the hadronic cross section fails for small hadronic sizes (large $\bar{Q}^{2}$), but is recovered for larger hadronic sizes.  The value of $x^{'}$ used in a calculation of the hadronic cross section will be significantly larger than $x$ for small hadronic sizes, leading to a suppression of the cross section in the small size region.  In particular, in Fig.~\ref{fig:dipoleg}, the approach to the BBL at small $d$ is slowed due to the large values of $x$ needed to push the small size configuration on shell.  In investigating the hadronic profile function, it may also be reasonable to determine $Q^{2}$ by letting it equal $\bar{Q}^{2}$ so that the value of 
 $d$ always corresponds to a typical component of the virtual photon.  We have done this in Fig.~\ref{fig:profile_9_23} and we can see that at small $d$,
 $\Gamma_{h}(b)$ is substantially larger, especially at small $x$.  Comparing Figs.~\ref{fig:dipoleg} and~\ref{fig:profile_9_23}, we see that at $d = .1$ fm this effect is significant while at intermediate hadronic sizes the effect is very small.  For $d \gtrsim .5$ fm there is no discernible difference between the two cases.  The physical meaning of this effect is that the profile function for a small size configuration approaches the BBL more slowly if it is far off shell for a given $Q^{2}$.  Note that once we begin to calculate the total cross section, an external value for $Q^{2}$ is explicit, and we no longer have this ambiguity.

Furthermore, there is some uncertainty in the gluon distribution used to calculate $\sigma_{tot}$.  This is demonstrated in Figs.~\ref{fig:F2vsx} and~\ref{fig:Flvsx} where we compare results for the structure functions using CTEQ5L~\cite{Lai:1999wy}, CTEQ6L~\cite{Pumplin:2002vw}, and MRST98~\cite{Martin:1998sq} leading order gluon parton distributions.  The dependence upon the parton distribution is seen to be small, but we used CTEQ5L parton distributions for all other calculations because they seem to yield optimal consistency with data.    

\begin{figure*}
\rotatebox{270}{\includegraphics[scale=0.60]{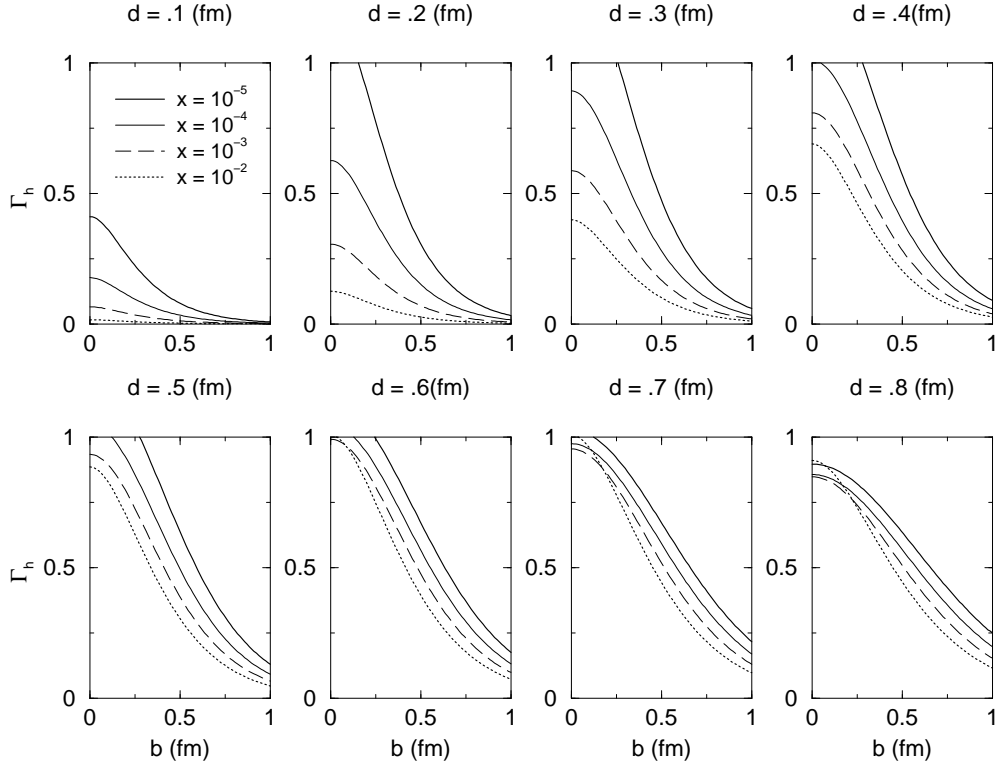}}
\caption{\label{fig:profilewrong}The hadronic configuration-nucleon profile function for different $x$ values.  Here we have used $d_{\pi} = .8$ fm.  Compare with Fig.~\ref{fig:dipoleg}. }
\end{figure*}

As we mentioned in Sect.~\ref{sec:model}, the value of $d_{\pi}$ that we used is consistent with the slope of the $\pi$N cross section as measured in Ref.~\cite{Burq:1981nv} and with the matching ansatz used in Ref.~\cite{McDermott:1999fa}.  
In the model of the $t$-dependence, $d_{\pi}$ determines where soft Pomeron behavior becomes important, and one may well ask whether a different value of $d_{\pi}$ is appropriate.  For models with a larger value of $d_{\pi}$, the suppression of the profile function due to the Pomeron form factor, $F_{P}$, does not occur until one considers larger hadronic configurations.  Therefore, for intermediate hadronic sizes, the profile function rapidly approaches the unitarity limit as $x$ decreases when $d_{\pi}$ is large.  This can be seen in Fig.~\ref{fig:profilewrong} where we have repeated the calculation of Fig.~\ref{fig:dipoleg}, this time using $d_{\pi} = .8$ fm.  Note the large values of $\Gamma_{h}(b)$ at small $b$ for $d \sim .5$ fm.  Therefore, by choosing a smaller value for $d_{\pi}$ we are making a conservative estimate of the approach to the BBL.     

\section{Estimating the Proximity of the Total $\gamma^{\ast} N$ Cross Section to the BBL}
\label{sec:gamma}

To properly study the proximity of DIS to the unitarity limit, we must evaluate the degree to which the different hadronic components contribute to the $\gamma^{\ast} N$ cross section, $\sigma^{ \gamma^{\ast} N}_{L,T}$.  $T$ and $L$ refer, respectively, to the transverse and longitudinal cross sections.  In the color dipole formalism, the 
 longitudinal and transverse cross sections can be factorized into the convolution of the perturbative 
light-cone wavefunction with a universal color dipole cross section,
\begin{eqnarray}
\sigma^{ \gamma^{\ast} N}_{L,T}(Q^{2},x) = \int_{0}^{1} dz \int d^{2} \vec{d}  \left| \psi_{L,T} (z,d) \right|^{2} \hat{\sigma}_{tot}(d,x^{'})\,. \label{eq:totalcross}
\end{eqnarray}
In this paper, Eq.~(\ref{eq:totalcross}) applies also to cases where $\hat{\sigma}_{tot}(d,x)$ is the cross section for interactions of large size hadronic configurations with the nucleon.
$z$ is the quark momentum fraction, and $\psi_{L,T}(z,d)$ is the longitudinal/transverse photon wave function calculated in QED.  We have calculated $\sigma^{\gamma^{\ast} N}_{L,T}$ using $\hat{\sigma}(d,x)$ with $t$-dependence determined in Sect.~\ref{sec:model}. 
Plots of $F_2$ and $F_L$ are shown in Fig.~\ref{fig:F2vsx} and Fig.~\ref{fig:Flvsx}.  The structure functions $F_2$ and $F_L$ are defined as,
\begin{eqnarray}
&&F_{L}(x,Q^{2}) = \frac{Q^{2}}{4 \pi^{2} \alpha_{e.m.} } \sigma_{L},  \nonumber \\ 
&&F_{2}(x,Q^{2}) = \frac{Q^{2}}{4 \pi^{2} \alpha_{e.m.} } ( \sigma_{L} + \sigma_{T} ) \,. 
\end{eqnarray}  
Note that the $t$-dependence is not needed for calculations of $F_2$ and $F_L$.  For more plots of the total cross section calculated within the QCD improved dipole model see Ref.~\cite{McDermott:1999fa}.  Figure~\ref{fig:dist} shows the distribution of the integrand in Eq.~(\ref{eq:totalcross}) over total hadronic sizes and demonstrates the suppression of large size hadronic configurations. 

\begin{figure*}
\rotatebox{270}{\includegraphics[scale=0.60]{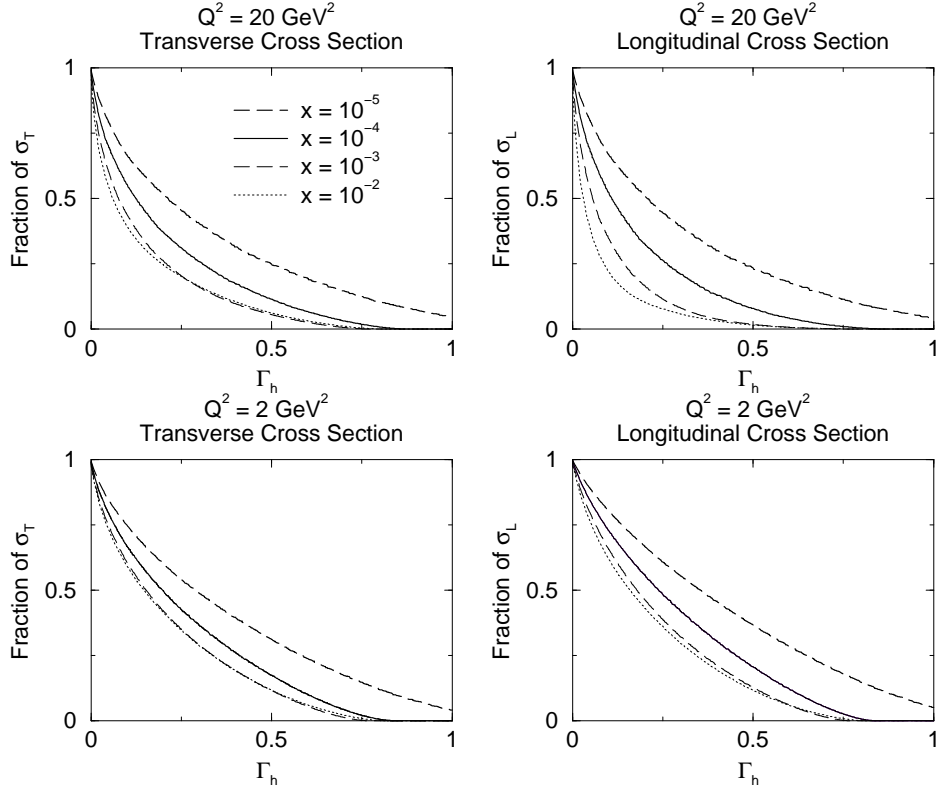}}
\caption{\label{fig:totalfrac}The fraction of $\sigma_{L,T}^{\gamma^{\ast}N}$ with contributions coming from values of $\Gamma_{h}$ greater than the corresponding values listed on the $x$-axis.  When $Q^{2} = 2$ GeV$^{2}$ and $x = 10^{-4}$, about 1/5 of $\sigma^{\gamma^{\ast} N}_{t\
ot}$ comes from hadronic components scattering with $\Gamma_{h} > 1/2$.}
\end{figure*}

We would like to study the contribution of intervals of $\Gamma(b)$ to the $\gamma^{\ast}N$ cross section.  However, the profile function for $\gamma^{\ast} N$ scattering by itself is not useful because the photon wave function is not normalizable and because it depends on $\alpha_{e.m.}$.  Thus, in order to look for the proximity to the BBL, we have plotted the fraction of the total $\gamma^{\ast}N$ cross section due to different regions of the \emph{hadronic} profile function.  Plots with different values of $x$ are shown in Fig.~\ref{fig:totalfrac}.  The $y$-axis denotes the fraction of the 
 longitudinal (transverse) cross section with contributions from $\Gamma_{h}(b)$ greater than the corresponding value on the $x$-axis.
 Note that when $Q^{2} = 2$ GeV$^{2}$ and $x \sim 10^{-4}$, about 1/5 of the longitudinal cross sections are due to dipole configurations corresponding to $\Gamma_{h}(b) > 1/2$.  

  At $Q^{2} = 20$ GeV$^{2}$, Fig.~\ref{fig:totalfrac} demonstrates the recovery of the leading twist behavior, especially for $\sigma_{L}$ at $x = .01$, where less than one-tenth of
 $\sigma_{L}$ is due to hadronic configurations with $\Gamma_{h}(b) > 1/2$.  
It is clear from Fig.~\ref{fig:totalfrac} that, for central impact parameters and low enough $Q^{2}$, a significant portion of the total cross section is due to hadronic configuration-nucleon interactions that are close to the unitarity limit.  In high energy $\gamma^{\ast}$A scattering, where the effects of the BBL are enhanced, we may be able to use DIS to probe the BBL.  This possibility is discussed within the context of the QCD improved dipole model in Sect.~\ref{sec:nuclear}.

\section{Comparison with Results of Other Studies}
\label{sec:compare}
\begin{figure}
\rotatebox{270}{\includegraphics[scale=0.60]{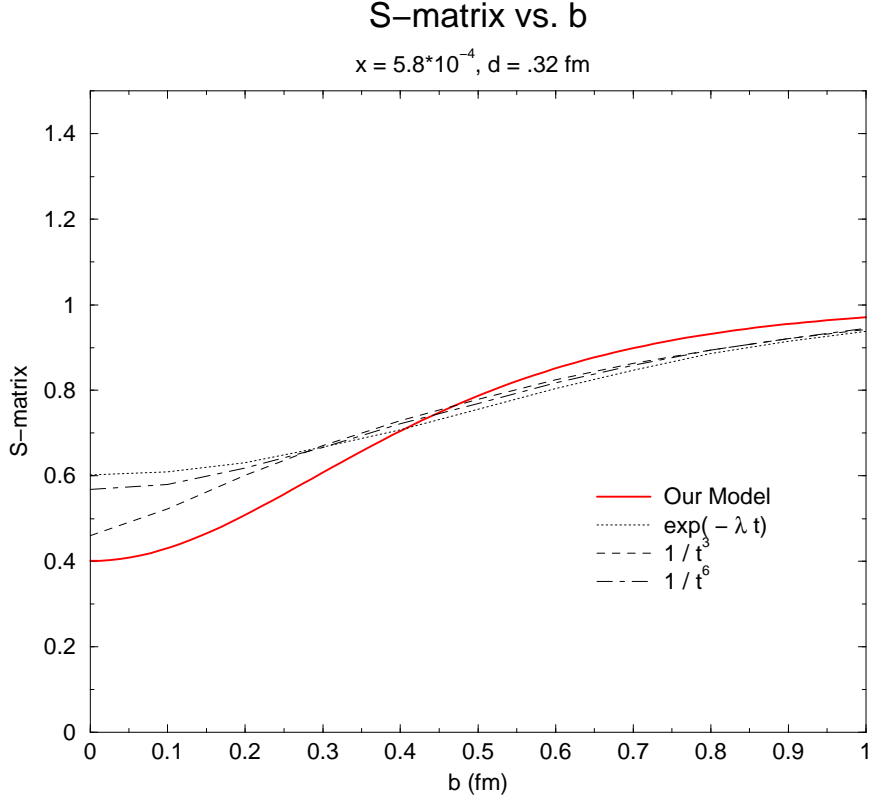}}
\caption{\label{fig:mueller}Comparison of the S-matrix calculated using Eq.~(\ref{eq:ourmodel}) with results obtained in \cite{Munier:2001nr}\
.  The bold line shows our result, while the dashed and dotted lines show results taken from Ref.~\cite{Munier:2001nr} using $\rho$ production with three different interpolations for the $t$-dependence.}
\end{figure}

The reasonableness of our model is demonstrated in Fig.~\ref{fig:F2vsx} where the dipole model is seen to be consistent with recent HERA data for $F_2$ at low values of $Q^{2}$ and $x$.  Furthermore, in Fig.~\ref{fig:Flvsx}, our model is seen to be consistent with preliminary results from HERA for $F_L$~\cite{Lobodzinska}.  
Other studies of the impact parameter picture of hadronic interactions with nucleons
 were done in \cite{Munier:2001nr}, where $\rho$ production data was used to extract the $t$-dependence.  The analysis in \cite{Munier:2001nr} used the S-matrix convention, $S(b) = 1 - \Gamma(b)$, in place of the profile function.  In Fig.~\ref{fig:mueller} we have plotted our prediction of the S-matrix profile for central impact parameters along with earlier result from Munier {\it et al.}~\cite{Munier:2001nr}.  In their analysis, the authors were restricted to using $\rho$ production data to model the $t$-dependence.  Data for $\rho$ production is limited to kinematics where $-t \lesssim 0.6$ GeV$^{2}$ ($b \gtrsim .3$ fm) so the accuracy of their results is limited to moderate impact parameters.  Further complication is introduced into their analysis by the need to model the $\rho$-meson wavefunction.  Our model uses $J/ \psi$ production data and is therefore valid at small values of $b$.  
As mentioned in Sect.~\ref{sec:model},
production of $J/ \psi$ depends only on the two-gluon form factor, and can therefore be extended down to very small impact parameters.  It is natural to compare our results for $S(b)$ with the median value of the dipole size as it is was used in the calculation of the amplitude estimated in Ref.~\cite{Frankfurt:1997fj}.  In our case, the value of $d$ that corresponds to $S(b)$ evaluated at $Q^{2} = 7$ GeV$^{2}$ is about $d \approx 0.32$ fm~\cite{thank}.  With this assumption, Figure ~\ref{fig:mueller} shows that our model has very good agreement with the results of Ref.~\cite{Munier:2001nr} at moderate values of $b$ while there is an expected deviation between the two models for low values of $b$ (see footnote \cite{consistent}).  Since our model will have $\sim t^{-4}$ behavior at small values of $b$, then even in the small $b$ region, our model deviates from the results of Ref.~\cite{Munier:2001nr} by no more than about twenty-five percent.  Reference \cite{Munier:2001nr} used a simple exponential or power ansatz to interpolate to larger values of $t$ as indicated in the figure.  In Fig.~\ref{fig:mueller}, it is seen that the model used in \cite{Munier:2001nr} has a high degree of uncertainty at small values of $b$ because of the necessity to guess the form of the function that interpolates to large $t$.  In contrast, our model uses information about $J/\psi$ production and DIS to model the small $b$ behavior.  

After preliminary results of our study were presented, there appeared
an experimental analysis with improved data on inclusive cross sections and vector meson production at HERA by Henri Kowalski and Derek Teaney \cite{Kowalski:2003hm}.  
They carry out an analysis similar to that used in \cite{Munier:2001nr}.  Therefore, it differs from our analysis in that
 it does not include information about large $t$ behavior of the two-gluon form factor.  Future improvements on the dipole picture should make comparisons with this data.

\section{Scattering off a Heavy Nuclear Target}
\label{sec:nuclear}
It is interesting to examine how the profile function $\Gamma_{h}(s,b,d)$ changes
when the free proton target is substituted by a heavy nuclear target such
as the nucleus of $^{208}$Pb. In the heavy nucleus case, the procedure for
obtaining $\Gamma_{h}(s,b,d)$  differs from the one in the nucleon case and is outlined below. First, for dipoles of small transverse sizes, $d < 0.2$ fm,
the inelastic scattering cross section at a given impact parameter $b$ 
is given by the perturbative QCD expression involving the impact parameter 
dependent nuclear gluon distribution, $g_A(x,Q^2,b)$ (compare to Eq.~(\ref{eq:pqcd}))
\begin{eqnarray}
\hat{\sigma}_{pQCD}^{inel}(d,x,b)=\frac{\pi^{2}}{3}d^{2} \alpha_{s} (\bar{Q}^{2})xg_A(x^{\prime},\bar{Q}^{2},b) \,,
 \label{eq:pb1} 
\end{eqnarray}
where $x^{\prime}$ is given by Eq.~(\ref{eq:xprime}).
The gluon distribution $g_A(x,Q^2,b)$, normalized such that
$\int d^2 b\, g_A(x,Q^2,b)=g_A(x,Q^2)$, was evaluated in 
\cite{Frankfurt:2002kd} using the theory of leading twist nuclear shadowing.
The profile function $\Gamma_{h}(s,b,d)$ can be found from Eq.~(\ref{eq:inelcross}) (see see also~\cite{Pumplin:na}),
\begin{equation}
2 Re \Gamma_{h}(s,b,d)-|\Gamma_{h}(s,b,d)|^2=\hat{\sigma}_{pQCD}^{inel}(d,x,b) \,,
\label{eq:pb2}
\end{equation}
Ignoring the small imaginary part of $\Gamma_{h}(s,b,d)$, which is even smaller  
  in the heavy 
nucleus case than in the free proton case because of the effect of nuclear
shadowing, Eq.~(\ref{eq:pb2}) gives
\begin{equation}
\Gamma_{h}(s,b,d)=1-\sqrt{1-\hat{\sigma}_{pQCD}^{inel}(d,x,b)} \,,
\label{eq:pb3}
\end{equation}
which is valid for $d < 0.2$ fm.
\begin{figure*}
{\includegraphics[scale=0.60]{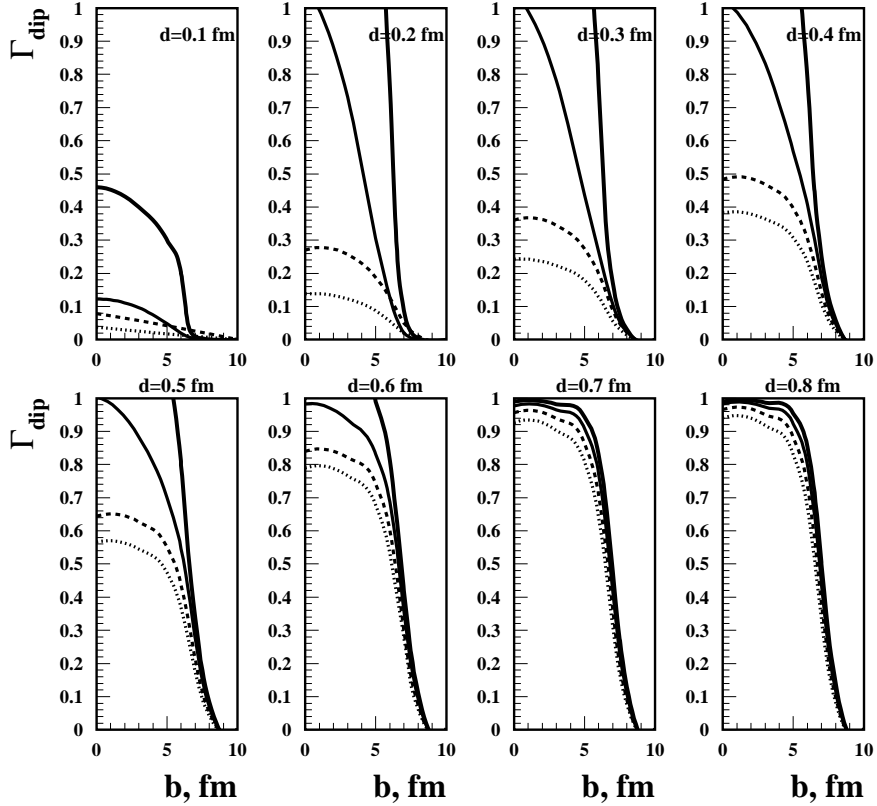}}
\caption{\label{fig:pb}The hadronic configuration-nucleus ($^{208}$Pb) profile function. The upper solid curves correspond to $x=10^{-5}$, and the lower solid curves corresponds to $x=10^{-4}$; the dashed curves correspond to $x=10^{-3}$; the dot-dashed curves correspond to $x=10^{-2}$.}
\end{figure*}
\begin{figure*}
\rotatebox{270}{{\includegraphics[scale=0.60]{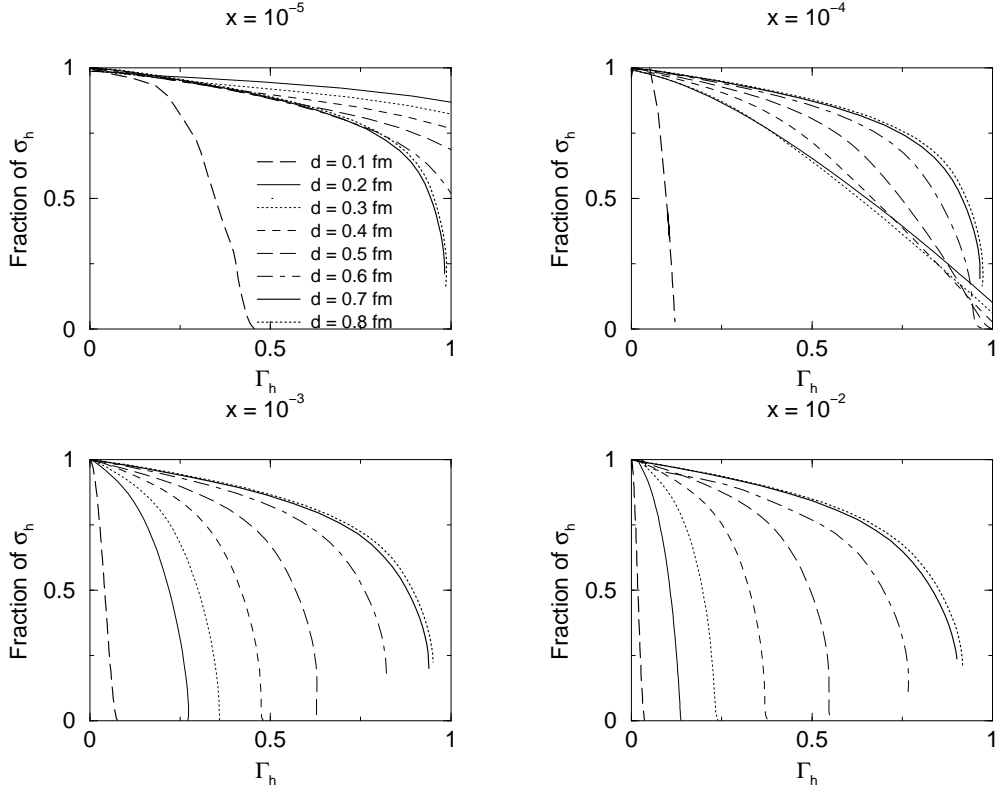}}}
\caption{\label{fig:nuclearcase}These plots are the analogue of those appearing in Fig.~\ref{fig:fraction}. They correspond to the profile functions for the nuclear target in Fig.~\ref{fig:pb}.  In these plots, $x^{'}$ corresponds to $Q^{2} = 2$ GeV$^{2}$.
}
\end{figure*}
\begin{figure*}
\rotatebox{270}{{\includegraphics[scale=0.60]{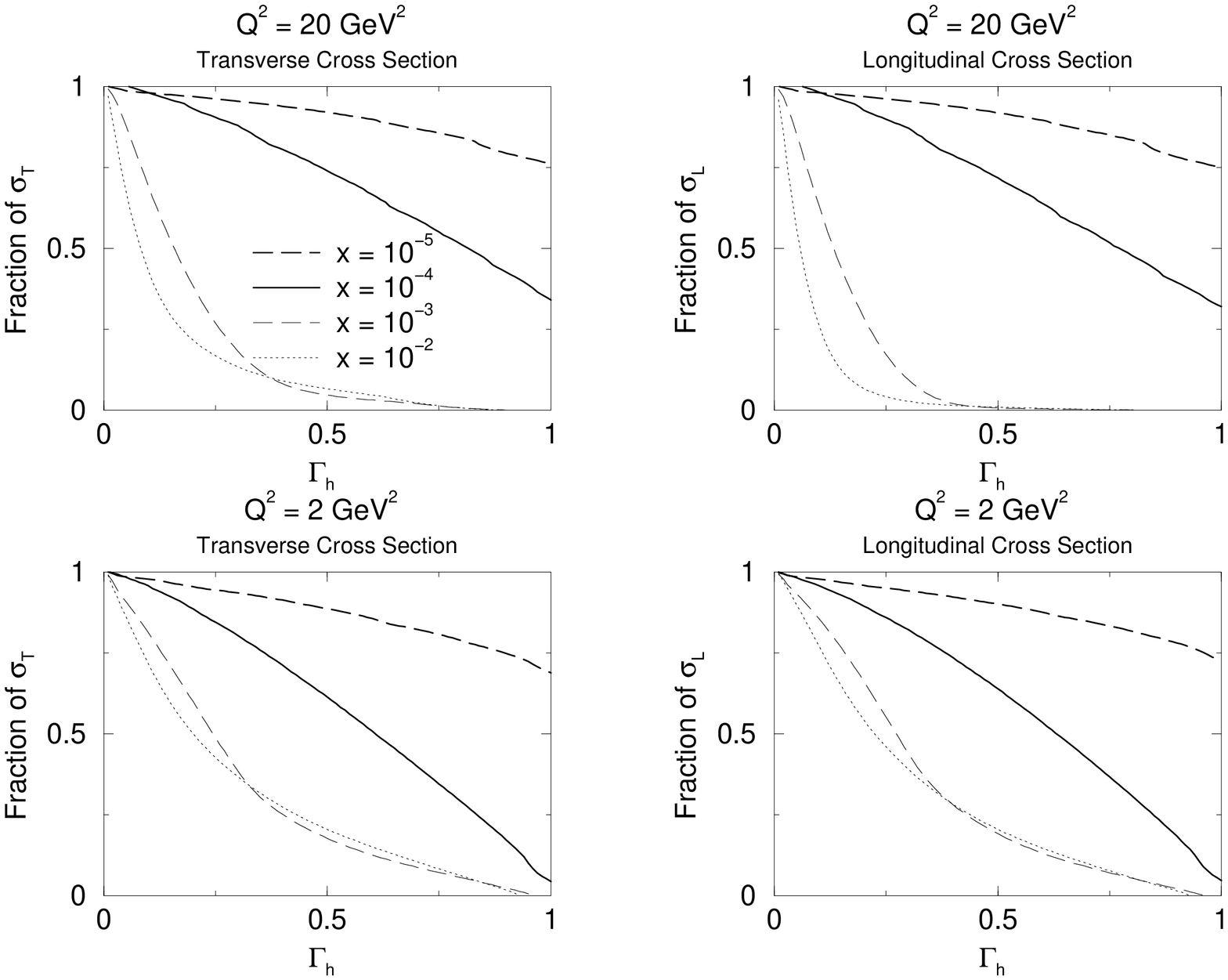}}}
\caption{\label{fig:nuclearcase2}These plots are the analogue of those appearing in Fig~\ref{fig:totalfrac}.  The fraction of $\sigma^{\gamma^{*} A}$ due to values of the hadronic profile function larger than $\Gamma_{h}$ is plotted versus $\Gamma_{h}$. 
}
\end{figure*}
Second, for dipoles of a larger size, $d_0=0.2 < d < d_{\pi}=0.65$ fm, the profile
function is found by interpolating between the pQCD expression of Eq.~(\ref{eq:pb3}) and the profile function calculated at $d=d_{\pi}$~\cite{McDermott:1999fa}
\begin{widetext}
\begin{equation}
\Gamma_{h}(s,b,d)=\left(\Gamma_{h}(s,b,d_{\pi})-\Gamma_{h}(s,b,d_0)\right) \frac{d^2-d_0^2}{d_{\pi}^2-d_0^2} + \Gamma_{h}(s,b,d_0) \,.
\label{eq:pb4}
\end{equation}
\end{widetext}
The profile function $\Gamma_{h}(s,b,d_{\pi})$ is calculated using the Glauber multiple scattering formalism~\cite{Glauber:1955qq}
\begin{equation}
\Gamma_{h}(s,b,d_{\pi})=1-e^{A \sigma_{\pi \, N}(s) T(b)/2} \,,
\label{eq:pbx}
\end{equation} 
where $\sigma_{\pi \, N}(s)$ is the energy-dependent pion-nucleon total scattering
cross section, $\sigma_{\pi \, N}(s)=23.78 (s/s_{0})^{0.08}$ mb, $s_{0} = 200$ GeV; $T(b)$ is the      
nuclear optical density normalized such that $\int d^2 b \, T(b)=1$.  $A$ is the number of nucleons in the target.  

Third, for the dipoles with the size $d > d_{\pi}$, the profile function is 
given by Eq.~(\ref{eq:pbx}), where the pion-nucleon cross section is allowed
to slowly grow as
\begin{equation}
\sigma_{\pi \, N}(s,d)=\sigma_{\pi \, N}(s) \frac{1.5 \,d^2}{d^2+d_{\pi}^2/2} \,.
\label{eq:pb5}
\end{equation}

The results for the profile function $\Gamma_{h}(s,b,d)$ for the nucleus of $^{208}$Pb are presented in Fig.~\ref{fig:pb} by the two solid curves ($x= 10^{-4}$ and $x= 10^{-5}$), dashed ($x=10^{-3}$) and dot-dashed ($x=0.01$) curves.  
 
The profile function for the nuclear target shows some similarity with the profile function for the proton target.  The main differences are that the BBL is approached over a larger range of impact parameters than in the case of a proton target.  This is not surprising because of the larger thickness of the nuclear target.  The plots in Fig.~\ref{fig:nuclearcase} show the fraction of the hadronic cross section due to large values of $\Gamma_{h}$.  

Large leading twist gluon shadowing tames the growth of the interaction of hadronic components of the photon with the nucleus so that the unitarity constraint is satisfied for $x \gtrsim 10^{-4}$ while the BBL may be reached for a large range of impact parameters.  For smaller $x$, unitarity starts to break down at central impact parameters.  For large $d$, unitarity is automatically satisfied since the Glauber model for large total cross sections leads to a $\Gamma_{h}$ that approaches unity.

Finally, we have included plots in Fig.~\ref{fig:nuclearcase2} for the nuclear target showing the fraction of $\sigma^{\gamma^{\ast} A}$ due to large values of $\Gamma_{h}$ analogous to those in Fig.~\ref{fig:totalfrac}.  In Fig.~\ref{fig:nuclearcase2} we see that the BBL is approached for nearly all values of $x$ at $Q^{2}=2$ GeV$^{2}$ and $Q^{2}=20$ GeV$^{2}$.  (Note the recovery of leading twist behavior for the longitudinal cross section at large $x$ and $Q^{2}=20$ GeV$^{2}$.)  Notice also that the fraction of $\sigma^{\gamma^{\ast}A}$ due to large values of $\Gamma_{h}$ for $x=.01$ is actually larger in some cases than for the case, $x=.001$.  This effect can be explained qualitatively by inspection of Fig.~\ref{fig:pb}.  For the case of the nuclear target, the main contributions to $\sigma^{\gamma^{\ast}A}$ come from smaller values of $d$ ($d \approx .2$ fm).  The growth of the profile function with decreasing $x$ at small $d$ is slower for smaller values of $x$ ($x \approx .01$) than for larger values.  Thus, the tail of profile function at large impact parameter may become significant in these regions.     
\section{Conclusion}

A general, well-known feature of $\gamma^{\ast}N$ and $\gamma^{\ast}A$ scattering is that the fraction of the interactions of hadronic components in the virtual photon wavefunction with the proton which take place at or near the BBL increases as $x$ and $Q^{2}$ decrease as is exhibited explicitly in this paper in Figs. \ref{fig:dipoleg}, \ref{fig:fraction} and \ref{fig:totalfrac}.  We hope that one day we may be able to exploit the novel properties of interactions in the BBL to study a new phase of pQCD.  In Ref.~\cite{Frankfurt:2001nt} the signatures of the BBL for DIS were discussed with the hopes that they would be seen in future experiments.  It remained to be determined, however, in which kinematical regions one can expect to see black body behavior.  Having now constructed a model of the amplitude for the interaction of the hadronic components of the virtual photon wavefunction, we are in a position to make rough estimates of the fraction of the hadronic interactions that exhibit the characteristic behavior of black body interactions.  More precisely, since we know that the effects that we have ignored so far -- inelastic diffraction and the real part of $A_{hN}$, -- will tend to increase the proximity of the interactions to the unitarity limit, then we can place lower limits on the values of $x$ and $b$ where a significant fraction of the events will occur at or near the BBL.  Our results show that, within available HERA kinematics, a significant fraction of the total DIS cross section is due to interactions of the hadronic components with the proton that occur near the BBL.  In particular, Fig.~\ref{fig:totalfrac} shows that at $Q^{2} \approx 2.0 $ GeV$^{2}$ and $x \lesssim 10^{-4}$, about 1/5 of the longitudinal cross section is due to values of $\Gamma_{h}(s,b,d) \gtrsim 1/2$.  The agreement of our model with preliminary HERA data and with previous models helps to strengthen this conclusion.  An improved model, with corrections for inelastic diffraction, will likely predict a more rapid approach to the BBL at small $x$ and central impact paramters.  The approach to the BBL as $x$ and $b$ decrease occurs much more rapidly for the case of a heavy nuclear target than for the case of a proton target.  This can be seen by comparing Figs. \ref{fig:fraction} and \ref{fig:nuclearcase}.  For example, at $d = 0.4$ fm and $x \approx 10^{-4}$, Fig.~\ref{fig:fraction} shows that, for the proton, less than 1/2 of the total cross section is due to contributions from $\Gamma_{h}(s,b,d) > 0.5$ whereas with a $^{208}$Pb target, Fig.~\ref{fig:nuclearcase} shows that over seventy percent of the total cross section is due to contributions from $\Gamma_{h}(s,b,d) > 0.5$.  This suggests that nuclear targets are ideal for studies of the BBL as a phase of QCD as has been discussed before in Ref.~\cite{Frankfurt:2001nt}.  Future work on this subject should incorporate inelastic effects.  Also, a greater understanding of the large $t$ behavior would lead to greater accuracy in the model.

\begin{acknowledgments}
We would like to thank J. Collins, L. Frankfurt, A. Mueller, and C. Weiss for useful discussions.  M. Strikman is an Alexander-von-Humboldt fellow.
\end{acknowledgments}

\end{document}